\documentclass[12pt]{article}
\pdfoutput=1

\usepackage[utf8]{inputenc}
\usepackage{times}
\usepackage{sectsty}
\sectionfont{\large}

\usepackage{fullpage}

\usepackage[parfill]{parskip}

\usepackage{amssymb,amsmath}

\usepackage[unicode=true]{hyperref}
\hypersetup{breaklinks=true, bookmarks=true, colorlinks=true}
\usepackage{authblk}

\title{\large \bf Nonparametric Bayesian grouping methods for spatial time-series data}

\author[1]{\normalsize Edward B. Baskerville}
\author[2]{\normalsize Trevor Bedford}
\author[3]{\normalsize Robert C. Reiner}
\author[1,4]{\normalsize Mercedes Pascual}
\affil[1]{\small Department of Ecology \& Evolutionary Biology, University of Michigan}
\affil[2]{\small Institute of Evolutionary Biology, University of Edinburgh}
\affil[3]{\small University of California, Davis; Fogarty International Center, National Institutes of Health}
\affil[4]{\small Howard Hughes Medical Institute}
\date{\normalsize 19 June 2013}

\newcommand{\bg}{\mathbf{g}}
\newcommand{\bP}{\mathbf{P}}
\newcommand{\bp}{\mathbf{p}}
\newcommand{\boldell}{\boldsymbol{\ell}}
\newcommand{\balpha}{\boldsymbol{\alpha}}
\newcommand{\bN}{\mathbf{N}}
\newcommand{\bn}{\mathbf{n}}
\newcommand{\Prob}{\mathrm{Prob}}
\newcommand{\like}{\mathcal{L}}
\newcommand{\argmax}{\mathrm{argmax}}
\newcommand{\mol}{\mathrm{mol}}

\begin{document}

\maketitle

\normalsize \begin{abstract}

\noindent We describe an approach for identifying groups of dynamically similar locations in spatial time-series data based on a simple Markov transition model. We give maximum-likelihood, empirical Bayes, and fully Bayesian formulations of the model, and describe exhaustive, greedy, and MCMC-based inference methods. The approach has been employed successfully in several studies to reveal meaningful relationships between environmental patterns and disease dynamics.

\end{abstract}

\section{Introduction}
\label{intro}

Spatial time series resulting from complex dynamical processes, for example environmentally driven disease dynamics, can present a formidable modeling challenge. The purpose of this document is to describe a conceptually simple approach for drawing insights from such data by identifying groups of spatial locations with similar dynamics. Dynamics are assumed to follow a simple nonparametric Markov transition model, thus eliminating the difficulty of formulating a mechanistic model.

Aspects of this approach have been employed in several studies of spatial disease dynamics. The basic Markov transition model is the core of a more complex model that was used to identify relationships between cholera and environmental variables in Dhaka, Bangladesh \cite{reiner2012}, but with groups imposed \emph{a priori} rather than inferred. The empirical Bayes formulation described here has been used to infer good groupings of regions for a study on malaria and irrigation in northeast Gujarat \cite{baeza2013}, as well as for several other projects in progress. This document does not detail any results, but instead serves as an extended description of the methods used in those studies as well as some extensions. The fully Bayesian approach is akin to one used for studying trophic food webs \cite{baskerville2011}.

In section \ref{data}, we describe the structure of the data. In section \ref{model}, we describe the Markov transition model and give maximum-likelihood, empirical-Bayes, and fully Bayesian formulations. In section \ref{inference}, we describe exhaustive, greedy, and Markov-chain Monte Carlo methods for inferring good groupings under the model.

\section{Data representation}
\label{data}

For these methods, we assume that the data set consists of time series for a discrete set of spatial regions. The model is based on a discretization of the data into a finite number of levels. How the discretization is performed implies particular assumptions about the system. For case data of an infectious disease, we suggest the following procedure as a sensible default:

1. If the data set consists of absolute counts, first normalize by the population over time within each spatial region in order to yield a prevalence time series (cases per capita) for each region.
2. Assign all zeros in the data to level $\ell = 0$.
3. Choose a desired number of levels $L$, and divide the remaining nonzero data into $L - 1$ equal quantiles. For example, if there are 10 nonzero data points and $L = 3$, the smallest five data points will have $\ell = 1$, and the largest five will have $\ell = 2$.

The methods that follow are based entirely on this discrete representation in both space and time.

\section{Nonparametric Markov transition model}
\label{model}

The model used here arises from two simple assumptions: (1) that the behavior of a time series can be characterized using transition probabilities between discrete levels of disease prevalence; and (2) that groups of spatial locations will obey the same transition probabilities.

\subsection{Transition probabilities}
\label{model-probs}

With $L$ different levels, there are $L^2$ different possible transitions from a state $i$ to a state $j$. Each of these transitions is assigned a different probability in each group of locations,
\begin{align}
	p_{ij}^{(g)}
\end{align}
where $g$ is the group identifier. With $K$ total groups, there are $K L^2$ different transition probabilities. Furthermore, all the transition probabilities from a particular level to other levels must sum to 1:
\begin{align}
	\sum_{j=1}^L p_{ij}^{(g)} = 1 \, ,
\end{align}
for every starting level $i$ and every group $g$. The transition probabilities for a particular group $g$ can be represented as a matrix, $\bP_g$, which has entries $p_{ij}^{(g)}$, and whose rows sum to 1.

\subsection{Grouping of spatial locations}
\label{model-grouping}

The $R$ different regions may be assigned to any number of groups $K \leq R$. The assignment into groups can be represented as a vector $\bg$:
\begin{align}
\bg = \begin{pmatrix}
  g_1 \\ \vdots \\ g_R
\end{pmatrix}
\end{align}
where $g_r$ is the group number that location $r$ is assigned to, $1 \leq g_r \leq K$. Because the group numbers are arbitrary, assignment vectors are often written in a canonical form known as a restricted growth function, which requires new groups to appear sequentially from left to right: $g_r \leq g_i + 1$ for all $i < r$. Mathematically, each arrangement into groups is known as a \emph{set partition}, which is uniquely defined by a single restricted growth function, but from here we refer to it simply as a \emph{grouping}.

\subsection{Model formulation}
\label{model-formulation}

Within a single location $r$, we have a series of discrete levels over time,
\begin{align}
	\boldell_r = \left\{ 
		\ell_r^{(0)}, \ldots, \ell_r^{(T)}
	\right\} \, .
\end{align}
for $T + 1$ time points. If location $r$ is in group $g_r$, then the probability of a particular transition in the data, $\ell_r^{(t - 1)} \rightarrow \ell_r^{(t)}$, is governed by the transition probability for group $g_r$, $p_{\ell_r^{(t-1)}\ell_r^{(t)}}^{(g_r)}$. Therefore, the probability of the entire time series, given initial condition $\ell_0$, group assignment $g_r$, and transition probabilities for the group, $\bP_{g_r}$, is
\begin{align}
	\Prob\left[\boldell_r | \ell_0, g_r, \bP_g\right]
	&= \prod_{t=1}^{T} p_{\ell_r^{(t-1)}\ell_r^{(t)}}^{(g_r)} \, .
\end{align}
This can be rewritten in terms of the number of transitions, $n_{ij}^{(r)}$, between each pair of levels $i$ and $j$ in location $r$:
\begin{align}
	\Prob\left[\bN_r | g_r, \bP_{g_r}\right]
	&= \prod_{i=0}^{L-1} \prod_{i=0}^{L-1} \left[p_{ij}^{(g_r)}\right]^{n_{ij}^{(r)}} \, ,
\end{align}
where $\bN$ is a matrix of all transition counts $n_{ij}^{(r)}$.

Since locations in the same group have the same transition probabilities, we can calculate the probability of the data associated with all locations in a single group according to $\bN_g$, a matrix of the total transition counts for an entire group $g$:
\begin{align}
	\Prob\left[\bN_g | \bg, \bP_{g}\right]
	&= \prod_{i=0}^{L-1} \prod_{i=0}^{L-1} \left[p_{ij}^{(g)}\right]^{n_{ij}^{(g)}} \, .
\end{align}
Note that $\bN_g$ is dependent on $\bg$, since it includes the counts for all locations in group $g$.

The probability of observing all time series in the data set is thus
\begin{align}
	\Prob\left[\left\{\boldell_1, \ldots, \boldell_m\right\} | \bg, 
		\left\{ \bP_1 \ldots \bP_K \right\} \right]
	&= \prod_{g=1}^K \prod_{i=1}^{L} \prod_{i=1}^{L}
		\left[p_{ij}^{(g)}\right]^{n_{ij}^{(g)}} \, .
\end{align}

\subsection{Maximum-likelihood inference}
\label{model-ml}

The probability of the data, given model parameters, can be rewritten as a function of the model parameters, given the data, a quantity known as the likelihood:
\begin{align}
	\like\left[\bg,  \left\{ \bP_1 \ldots \bP_K \right\} | \left\{\boldell_1, \ldots, \boldell_m\right\} \right]
	&=
	\Prob\left[\left\{\boldell_1, \ldots, \boldell_m\right\} | \bg,  \left\{ \bP_1 \ldots \bP_K \right\} \right]
\end{align}

Under maximum likelihood, the best estimate of the parameters is where the likelihood---the probability of observing the data---is maximized:
\begin{align}
	\label{eq:mle}
	\hat{\bg}, \left\{ \hat{\bP}_1 \ldots \hat{\bP}_K \right\} &= \argmax \, \like\left[\bg,  \left\{ \bP_1 \ldots \bP_K \right\} | \left\{\boldell_1, \ldots, \boldell_m\right\} \right] \, .
\end{align}

In other words, under maximum likelihood, assuming a fixed number of groups $K$ and a fixed discretization of the data, the goal is to find the combination of grouping and transition probabilities that makes the data most probable.

With the grouping $\bg$ specified, the maximum-likelihood estimate of each $p_{ij}^{(g)}$ is simply the empirical fraction of transitions for group $g$ that starting at level $i$ that ended up at level $j$:
\begin{align}
	\hat{p}_{ij}^{(g)} &= \frac{n_{ij}^{(g)}}{\sum_{j'} n_{ij'}^{(g)}} \, .
\end{align}

The best grouping, similarly, is the one that maximizes the likelihood under the maximum-likelihood estimates of the corresponding transition matrices. Methods for identifying the best grouping are discussed in section \ref{inference}.

\subsection{Bayesian formulation}
\label{model-bayes}

In a Bayesian model, the object to be inferred is a probability distribution over model parameters, rather than a single point estimate. Parameters that are deemed more likely, given the data, are given more weight in this distribution, called the \emph{posterior distribution} because it is the distribution of parameters given the data---that is, after the data has been observed. This requires assuming a \emph{prior distribution} over model parameters---the distribution before the data has been observed.

A fully Bayesian treatment can be built incrementally: first by introducing priors for transition probabilities, but identifying only a single best grouping; and then by introducing priors for the assignment into groups.

\subsubsection{Priors for transition probabilities}
\label{model-prob-priors}

Each row $\bn_i^{(g)}$ of the transition count matrix $\bN_i^{(g)}$ for group $g$ is governed by a categorical random variable with parameters from the corresponding row of the probability matrix, $\bp_i^{(g)}$.
\begin{align}
	\Prob\left[\bn_i^{(g)} | \bg, \bp_i^{(g)}\right]
&= \prod_{j=0}^{L-1} \left[p_{ij}^{(g)}\right]^{n_{ij}^{(g)}} \, .
\end{align}

The conjugate prior for the transition probabilities is the Dirichlet distribution, which has density
\begin{align}
	f(\bp_i | \balpha_i) &=
		\frac{
			\Gamma \left( \sum_{j=0}^{L-1} \alpha_{ij} \right)
		}{
			\prod_{j=0}^{L-1} \Gamma(\alpha_{ij})
		}
		\prod_{j=0}^{L-1}  \left[p_{ij}^{(g)}\right]^{(\alpha_{ij} - 1)}
\end{align}
where $\balpha_i$ is a vector of ``concentration parameters'' that governs the relative weight and evenness of the transition probabilities, and $\Gamma(\centerdot)$ is the gamma function.

This prior distribution is \emph{conjugate} because the posterior distribution is also Dirichlet, so that the parameters $\bp_i^{(g)}$ are governed by posterior concentration parameters $\alpha_{ij}^{(g)}$ equal to
\begin{align}
	\alpha_{ij}^{(g)} &= \alpha_{ij} + n_{ij}^{(g)} \, .
\end{align}

\subsubsection{Identification of prior concentration parameters}
\label{model-conc-priors}

The concentration parameters $\balpha_i$ for each starting level $i$ must be chosen in order to fully specify the model. Prior information about the system can be used to inform this choice, but if prior information is absent, one natural choice is the non-informative Jeffreys prior, with all $\alpha_{ij} = \frac{1}{2}$ \cite{jeffreys1946}.

Another approach is to apply the maximum-likelihood principle to the concentration parameters, while allowing transition probabilities to vary, an approach known as \emph{maximum marginal likelihood} or \emph{empirical Bayes} \cite{robbins1956}. If the grouping $\bg$ is fixed, this requires finding values of each $\alpha_{i}$ that maximize the probability of observing the data, integrated over all transition probabilities:
\begin{align}
	\hat{\balpha}_i &= \argmax_{\balpha_i} \prod_{g=1}^K \Prob \left[ \bn_i^{(g)} | \balpha_i \right] \, ,
\end{align}
where
\begin{align}
	\Prob \left[ \bn_i^{(g)} | \balpha_i \right] &= \int_{\bp_i^{(g)}} f(\bp_i^{(g)} | \balpha_i) \Prob\left[ \bn_i^{(g)} | \bp_i^{(g)} \right] \, d\bp_i^{(g)}
\end{align}
is analytically tractable for the conjugate prior, and equal to
\begin{align}
	\label{eq:marginal-row}
	\Prob \left[ \bn_i^{(g)} | \balpha_i \right] &=
	\frac{
		\Gamma \left( \sum_{j=0}^{L - 1} \alpha_{ij} \right)
	}{
		\Gamma \left( \sum_{j=0}^{L - 1} n_{ij}^{(g)} + \alpha_{ij} \right)
	}
	\prod_{j=0}^{L-1}
	\frac{
		\Gamma \left( n_{ij}^{(g)} + \alpha_{ij} \right)
	}{
		\Gamma \left( \alpha_{ij} \right)
	} \, .
\end{align}
This can be simplified by requiring that all $\balpha_i$ vectors have the same components, or further that all $\alpha_{ij}$ parameters, take on the same value.

\subsubsection{Identification of a single best grouping}
\label{one-grouping}

Rather than inferring a full posterior distribution over both transition probabilities and groupings, the empirical-Bayes approach can be used to identify a single best grouping while allowing for a distribution over transition probabilities. This can be done in conjunction with empirical-Bayes identification of the concentration parameters, but for simplicity we assume here that the concentration parameters $\balpha_i$ are fixed. The probability of observing the transition counts in all groups, given the grouping $\bg$, is:
\begin{align}
	\label{eq:marginal-grouping}
	\Prob
	\left[
		\left\{ \bN_1 \ldots \bN_K \right\} | \bg,
			\left\{ \balpha_0 \ldots \balpha_{L-1} \right\}
	\right]
	&=
	\prod_{g=1}^{K} \prod_{i=0}^{L-1}
	\Prob \left[ \bn_i^{(g)} | \bg, \left\{ \balpha_0 \ldots \balpha_{L-1} \right\} \right]
	\, ,
\end{align}
where the inner probability of a single row of transitions in a single group is as given in equation \eqref{eq:marginal-row}.

Via maximum likelihood, the best grouping is as given in equation \eqref{eq:mle}.

In section \ref{inference} we describe three different methods for searching through the space of possible groupings.

\subsubsection{Priors for groupings}
\label{model-group-priors}

The simplest approach in choosing a prior for groupings is to use a uniform distribution, so that all groupings are \emph{a priori} equally likely. This choice results in a posterior distribution over groupings that is simply proportional to equation \eqref{eq:marginal-grouping}. However, such a prior implies a distribution that is strongly biased toward an intermediate number of groupings.

Another approach is to use a combined Dirichlet-categorical distribution for groupings. Under this model, each region has some probability of being a member of each group $g$. The vector of group probabilities can in turn be modeled as being drawn from a Dirichlet distribution with a shared concentration parameter, since all groups are \emph{a priori} equivalent. As seen above when marginalizing over transition probabilities, the Dirichlet and categorical distributions can be analytically collapsed, yielding the following probability mass function dependent on shared concentration parameter $\delta$:
\begin{align}
	\Prob \left[ \bg | \delta \right] &=
	\frac{K!}{(K-d)!}
	\frac{
		\Gamma \left( K\delta \right)
	}{
		\Gamma \left( K\delta + \sum_{g=1}^{K} n_g \right)
	}
	\prod_{g=1}^{K}
	\frac{
		\Gamma \left( n_g + \delta \right)
	}{
		\Gamma \left( \delta \right)
	} \, ,
\end{align}
where $d$ is the number of nonempty groups and $n_g$ is the number of regions assigned to group $g$. The leading factor $\frac{K!}{(K-d)!}$ is required because empty groups are indistinguishable.

Another possible prior is the Dirichlet process, which can be seen as a limiting case of the Dirichlet-categorical model \cite{green2001} where the number of groups $K \rightarrow \infty$ and the concentration parameter $\delta \rightarrow 0$ such that $K \delta \rightarrow \alpha > 0$. The probability mass function of the Dirichlet process is
\begin{align}
	\Prob \left[ \bg | \alpha \right]
	&=
	\frac{
		\alpha^d \Gamma(\alpha) \prod_{g=1}^{d} (n_g - 1)!
	}{
		\Gamma(\alpha + R)
	} \, ,
\end{align}
where $R$ is the total number of regions being grouped, $d$ is the number of nonempty groups, and $n_g$ is the number of regions in group $g$.

\section{Inference and search methods for groupings}
\label{inference}

Depending on the nature of the data and the modeling approach, several methods are available for searching through the space of possible groupings. If the number of possible groupings of regions is small, then it may be possible to exhaustively enumerate all of them. When the combinatorics become intractable, approximation methods are required. When using maximum likelihood or empirical Bayes for group arrangements, a greedy search algorithm should be able to identify good groupings, but will fail to exhaustively explore the space. If a fully Bayesian treatment is desired, a Markov-chain Monte Carlo approach is necessary. Under maximum likelihood or empirical Bayes, stochastic optimization algorithms can be used as a faster alternative.

Identification of the maximum-likelihood grouping (equation \eqref{eq:mle}) makes sense as long as the number of parameters stays the same, which means that the number of nonempty groups must remain fixed. If different numbers of groups are being considered, then the comparison between groupings must penalize for the change in degrees of freedom, for example using the Akaike information criterion \cite{akaike1974}. However, unlike the Bayesian approach, which directly integrates over the uncertainty in parameter values, AIC penalizes all parameters equally based on assumptions of independence and asymptotic distributions. Therefore, the Bayesian approach is likely to be more robust.

\subsection{Exhaustive enumeration}
\label{enumeration}

The number of different possible assignments of regions to groups is equal to the Bell number $B_R$ for $R$ regions:
\begin{align}
B_R
&=
\sum_{r=0}^R \frac{1}{r!}\sum_{j=0}^{r}(-1)^{r-j} \binom{r}{j} j^R \, .
\end{align}

If this number is small, then good group arrangements can be found by enumerating every single arrangement, which might be reasonable for as many as 15 regions ($B_{15} \approx 1.4 \times 10^9$), but quickly grows out of reach ($B_{30} \approx 8.5 \times 10^{23} \approx 1.4 \mol$). Restricting the number of groups also increases the practical ceiling for exhaustive search. For example, when splitting into only two groups, 30 regions can be split only $2^{29} \approx 5.4 \times 10^8$ different ways, which is quite tractable on modern computers.

Note that exhaustive enumeration can be used not only to identify the best grouping under maximum likelihood or empirical Bayes, but can also be used to calculate the full posterior distribution over groupings given a prior (section \ref{model-group-priors}).

Because the likelihood and marginal likelihood calculations involve finding a sum of transition counts over all members of the group, enumeration can be vastly sped up by ensuring that each successive grouping is different from the previous one by a single change in group membership. That way, the transition counts can be updated by subtracting one integer per matrix entry from the old group, and adding one integer per matrix entry to the new group. This optimization can be achieved through the use of a \emph{Gray code}, which was originally invented to eliminate the need for synchronization of multiple electromechanical switches \cite{gray1953}. For example, a Gray code for five regions being organized into two groups is:
\begin{verbatim}
00000, 00001, 00011, 00010, 00110, 00111, 00101, 00100,
01100, 01101, 01111, 01110, 01010, 01011, 01001, 01000
\end{verbatim}
where the group numbers are 0 and 1. Note that the first group assignment is always 0, so that these groupings are in restricted growth function form.

\subsection{Greedy clustering}
\label{greedy}

A simple greedy heuristic is agglomerative clustering, which starts by putting every region in its own group. At each step, all possible combinations of two groups are tested, and the one that gives the greatest increase in marginal likelihood or AIC is selected. The process terminates either when all regions are in the same group or no increase in marginal likelihood is possible. This method is very fast, but does not search all possible clusterings and may arrive at one that is suboptimal.

\subsection{Markov-chain Monte Carlo using Gibbs sampling and MCMCMC}
\label{mcmc}

When exhaustive enumeration is computationally intractable, a stochastic search method must be used. Markov-chain Monte Carlo (MCMC) approaches, originally invented for sampling from distributions of the energy state of interacting molecules \cite{metropolis1953}, have the advantage of converging to a desired target distribution, and can thus give samples from the full Bayesian posterior over groupings and model parameters.

\subsubsection{Gibbs sampling of groupings}
\label{gibbs}

The most straightforward approach for this model is to use Gibbs sampling to move regions from group to group. Under Gibbs sampling, a new value for a parameter is drawn from the distribution conditional on all other parameters ; the sequence of samples will converge to the target posterior distribution \cite{geman1984}. Specifically, for this model, the algorithm repeats the following steps until convergence:

\begin{enumerate}
\item Randomly sample initial group assignments.
\item Repeat for each region $r$:
	\begin{enumerate}
	\item Calculate the marginal likelihood of the model with $r$ assigned to each of the possible groups, keeping all other regions' group assignments the same.
	\item Choose a new group for $r$ by sampling from the discrete probability distribution weighted by the marginal likelihoods.
	\end{enumerate}
\end{enumerate}

\subsubsection{MCMCMC}
\label{mc3}

A technique known as Metropolis-coupled Markov chain Monte Carlo (MCMCMC, or (MC)\textsuperscript{3})---can be used to avoid getting stuck at local maxima in the search space \cite{geyer1991}. Multiple chains are run at different ``heats,'' where hotter chains explore configurations more freely, and colder chains are more likely to move toward better solutions. This enables good solutions to be refined while other alternatives may still be explored.

A chain at heat level $i$ explores the distribution
\begin{align}
	f_i(\theta | D) &\propto f(\theta) \left[f\left(D | \theta \right) \right]^{\tau_i}
\end{align}
where $\theta$ represents parameters (e.g., groupings); $D$ is the data, and $0 \leq \tau_i \leq 1$.

Swaps are proposed between adjacent chains $i, j$, and the probability of accepting a swap between chains is equal to
\begin{align}
\Prob[\theta_i \rightarrow \theta_j, \theta_j \rightarrow \theta_i] &= \min \left\{ 
	1,
	\frac{
		f_i(\theta_j | D) f_j(\theta_i | D)
	}
	{
		f_i(\theta_i | D) f_j(\theta_j | D)
	}
\right\} \\
&=
\min \left\{ 
	1,
	\left[ \frac{
		f(D | \theta_j)
	}
	{
		f(D | \theta_i)
	} \right]^{(\tau_i - \tau_j)}
\right\}
\end{align}
where $\theta_i$ is the initial configuration of chain $i$ and $\theta_j$ is the initial configuration of chain $j$. The cold chain, with $\tau_0 = 0$, will converge to the target distribution $f(\theta | D)$.

\bibliographystyle{plos2009}
\bibliography{references}

\end{document}